\documentclass[mathleft
]{an}
\usepackage{graphicx}
\usepackage{times}
\usepackage{amssymb}
\begin{document}

\Pagespan{292}{}
\Yearpublication{2007}%
\Yearsubmission{2006}%
\Month{10}%
\Volume{328}%
\Issue{3/4}%
\DOI{10.1002/asna.200610732}

\title{Surface magnetic field effects in local helioseismology}

\author{H. Schunker\inst{1,2}\fnmsep\thanks{Corresponding author:
  {schunker@mps.mpg.de}\newline}
\and  D.C. Braun\inst{3}
\and{P.S. Cally\inst{1}
}
\institute{
Centre for Stellar and Planetary Astrophysics, Monash University, Melbourne, 
Australia 3800
\and 
Max-Planck-Institut fur Sonnensystemforschung, D-31791 Katlenburg-Lindau, Germany
\and 
NorthWest Research Associates, CoRA Division, 3380 Mitchell Lane, 
Boulder, CO 80301, U.S.A.}}
\titlerunning{Surface magnetic field effects}
\authorrunning{H. Schunker, D.C. Braun \& P.S. Cally}
\received{2006 Oct 20}
\accepted{2007 Jan 2}
\publonline{2007 Feb 28}

\keywords{Sun: helioseismology -- Sun: magnetic fields -- sunspots}

\abstract{%
Using helioseismic holography strong evidence is presented that the phase (or
equivalent travel-time) of
helioseismic signatures in Dopplergrams within sunspots depend upon the 
line-of-sight angle in the plane containing the magnetic field and vertical directions. This
is shown for the velocity signal in the penumbrae of two sunspots at 3, 4 and 5 mHz.
Phase-sensitive holography demonstrates that they 
are significantly affected in a strong, moderately inclined magnetic field.  
This research indicates that the effects of the surface magnetic field are potentially 
very significant for local helioseismic analysis of active regions. }

\maketitle

\section{Introduction}
The potential influence of the magnetic field at the surface on acoustic waves is 
somewhat controversial, but there is a growing consensus that surface magnetic effects
should be considered and included in local helioseismic analysis of active regions 
(Braun 1997; Woodard 1997; Zhao \& Kosovichev 2006; Braun \& Birch 2006).  The showerglass effect 
(Lindsey \& Braun 2005a,b) is a recently 
suggested phenomenon,  consisting of large amplitude and phase distortions 
of the surface wavefield due to magnetic fields in the photosphere. 
Lindsey \& Braun (2005b) apply a correction for these effects based on the strong 
correlation between magnetic field strength and the phase perturbations. A 
peculiar enhancement of the phase perturbation is noted in the penumbra of 
sunspots, called the penumbral acoustic anomaly (Lindsey \& Braun 2005a). The magnetic 
field in the penumbra is not as strong as in the umbra and this research 
attempts to quantify the phase perturbations produced specifically by 
inclined magnetic fields which characterize sunspot penumbrae.

This paper expands upon previous research by Schunker et al. (2005) and explores the effects 
that the inclined field within the penumbrae of two sunspots may have on 
acoustic waves originating below the solar surface. 
The analysis uses helioseismic holography (Sect. \ref{hol}). 
Generally, in the quiet Sun, the ``ingression'' (the deduced amplitude
of incoming acoustic waves from a surrounding pupil)  and the observed surface signal 
agree well (Lindsey \& Braun 2005a,b). In magnetic regions, a deviation 
of the amplitude and phase of the incoming acoustic waves is indicated. 
Schunker et al. (2005) demonstrated that 
there is a clear cyclic variation of the ingression phase with 
azimuthal angle within a sunspot penumbra, and the line-of-sight direction.
It is suggested that the effect may be due to the 
process of mode conversion as discussed by Schunker \& Cally (2006). It has also previously 
been established that mode conversion is able to simulate the observed acoustic 
absorption in inclined magnetic fields (Cally \& Crouch 2003; Cally, Crouch \& Braun 2005). This has 
encouraging prospects for explaining the observed penumbral and inclined field 
dependencies. Initially, a fast acoustic wave propagates up to the surface 
where it encounters the $a=c$ layer (where $a$ is the Alfv\'en speed and $c$ is 
the sound speed) and undergoes \textit{conversion} to a fast (magnetic) wave 
and \textit{transmission} to a slow (acoustic) wave. The amount of energy 
devoted to each depends on the attack angle between the path of the ray and the 
magnetic field where the conversion/transmission occurs (where $a=c$). It is 
thought that the slow acoustic wave is what may be contributing to the observed 
effect of Schunker et al. (2005), and that presented here 
(see also Schunker \& Cally 2006; Cally 2007).

The sunspot analysis is done using full disk data from the Michelson Doppler 
Imager (MDI) (Scherrer et al. 1995) over a period of days as the sunspot crosses the 
solar disk. The data is tracked and Postel 
projected into data cubes for each 
day. This enables the penumbral magnetic field to be viewed from a varying 
line-of-sight at different heliocentric angles (one for each day of 
observation). The phase shift of the incoming acoustic waves is determined from 
the correlation of the incident acoustic wave (estimated using holography) and 
the surface velocity.  Vector magnetograms obtained from the Imaging Vector 
Magnetograph (IVM) (Mickey et al. 1996), for both sunspots are used to 
determine the orientation and strength of the field in relation to the phase 
perturbation.

\section{Holography}\label{hol}

Based on observed surface signals helioseismic holography is able to detect 
subsurface wave speed variations (Lindsey \& Braun 2000). The amplitudes  of acoustic 
waves propagating through the interior are inferred at a focal point of 
particular depth, $z$, and horizontal position, $\vec {r}$. 

For this purpose the focal plane is located at the surface, $z=0$. The inferred contribution of the observed waves is calculated through the
interior and back up to the surface. The ingression is the superposition of the 
amplitudes of the incoming waves at $(\vec{r},z)$ and time $t$,
\begin{eqnarray} 
\lefteqn{H_{-} ( \vec{r}, z, t) = \int {{\rm d} t'} \int_{a < | \vec{r}- \vec{r}' | < 
b}   {\rm d}^{2} \vec{r}'  }\nonumber\\ 
&& \ \ \ \ \ \ \ \ \ \ \ \ \ \ \ G_{-}(| \vec{r}- \vec{r}'|, z, t-t')  \psi(\vec{r}', t), \ \ \ \ \ \ \ 
\  \ 
\label{ingress}
\end{eqnarray}
which is calculated here at $z=0$.  The Green's function $G_{-}$  represents  
the sub-surface  disturbance at ($\vec {r}, z, t$) resulting from  a unit 
acoustic impulse originating at surface co-ordinates ($\vec {r}', 0,  t'$). A 
Green's function derived in a wave-mechanical formalism that  includes effects 
for dispersion (Lindsey \& Braun 2000), is used here. The computation is confined to  an 
annulus or `pupil' surrounding the focal point $\vec {r}$ with inner and  
outer radius $a$ and $b$ respectively. Essentially, the ingression is  
what \textit{should} result  from the incoming acoustic waves  
propagating from the pupil to the focal point in the absence of  perturbations.

The ingression at the focus, $\vec{r}$, is correlated with the surface 
velocity signal at $\vec{r}$ to 
gauge the effect of any surface anomaly. This ``local ingression control correlation"  
(Lindsey \& Braun 2005a)  is given as
\begin{equation}
C( \nu) =  \langle \hat{H}_{-}(\vec{r}, \nu) \hat{\psi}^* (\vec{r}, 
\nu)\rangle_{\Delta \nu},
\label{correlation}
\end{equation}
where $\hat{\psi}$ represents the temporal Fourier transform of the surface 
disturbance $\psi$, $\nu$ is the temporal frequency, and $\hat{H}_{-}$ is the 
temporal Fourier transform of the ingression. 
The asterisk denotes complex conjugation and the brackets indicate an average 
over a positive frequency range  $\Delta \nu$ of 1 mHz centered at a certain 
frequency (here at either 3, 4 or 5 mHz).   The effect of the surface 
perturbations is quantified by the phase of the local ingression
correlation,
\begin{equation}
\delta \phi = \arg [ C( \nu) ].
\label{corr_phase}
\end{equation}

We employ observations of two mature, nearly axisymmetric sunspots.
The observations of the sunspot in AR9026 span a time from 2000 June 3 to June 
12, and the sunspot in AR9057 from 2000 June 24 to 2000 July 2. This allows 
 different line-of-sight observations of the sunspots as they rotate from 
approximately $60^\circ$ East of the central meridian to about $60^\circ$ West 
of the central meridian. These sunspots were selected as they do not evolve 
significantly over the duration of the observations.

As Schunker et al. (2005) described, the full-disk MDI data were analyzed in 24 hour 
sets. For each day Postel  projections were made, centered near the sunspots. 
From the temporal Fourier transform, the ingression was computed and correlated 
to the surface Doppler signal (Eqs. \ref{ingress} and \ref{corr_phase}). 
The pupil size is $a=20.7$ Mm and $b=43.5$ Mm for the inner and outer radii 
respectively. At a frequency of 5 mHz, this pupil selects $p$-modes with 
spherical harmonic degrees ($\ell$) and radial order ($n$) between $\ell 
\approx$ 450 ($n$ = 5) and $\ell \approx$ 700 ($n$ = 4). This pupil is also of 
a size to eliminate the majority of the active region signal when computing the 
correlation inside the penumbra. Simple analysis of the acoustic dispersion 
relation near the surface shows that $\cot\eta\approx 2 \pi \nu R_\odot/c 
\ell$, where $R_\odot$ is the solar radius, $\eta$ is the propagation angle 
from vertical and $c$  is the local sound speed. At 3, 4 and 5 mHz, and in the 
given range of $\ell$, $\cot\eta\approx$ 10, indicating a primarily vertical 
propagation. 

\section{Ingression correlation phase} \label{results}

\begin{figure}
\begin{center}
\includegraphics*[width=82mm,height=82mm]{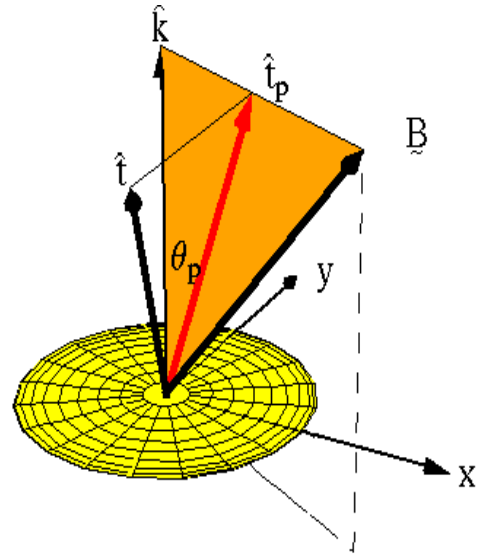}
\vspace{-3mm}
\caption{A representation of the geometry of $\theta_{\rm p}$. The `sunspot' has a 
radial vector $\hat{\vec{k}}$ and a magnetic field vector,  $\vec{B}$, 
creating a plane. The line-of-sight vector,  $\hat{\vec{t}}$, is projected 
onto this plane creating,  $\hat{\vec{t}_{\rm p}}$, which makes an angle 
$\theta_{\rm p}$ to $\hat{\vec{k}}$. This is essentially an independent parameter 
representing the line-of-sight viewing angle of the magnetic field. }
\label{theta_p}
\end{center}
\end{figure}

The penumbral regions for each sunspot, inner and outer radii, are determined 
from MDI continuum images as $ 7.3 - 16$ Mm from the centre of the sunspot in 
AR9026 and $6 - 13$ Mm for the smaller sunspot in AR9057. To determine the 
dependence of the phase shift (the ingression correlation phase) $\delta \phi$, 
on the line-of-sight direction projected onto the plane of the local
magnetic field, a parameter $\theta_{\rm p}$ is defined (see Fig.\,\ref{theta_p}). 
The line of sight vector, $\hat{\vec{t}}$, is projected onto the plane 
containing the radial vector, $\hat{\vec{k}}$, and the magnetic field vector 
$\vec{B}$. The projected line-of-sight vector, 
$\hat{\vec{t}_{\rm p}}$, thus makes an angle $\theta_{\rm p}$ with $\hat{\vec{k}}$.

The magnetic field vector is determined from IVM magnetogram data.  This 
provides the orientation and strength of the surface magnetic field in the 
sunspot. The IVM instrument observes in the same line as the MDI (Ni 676.8 nm), 
and hence at the same geometrical height in the photosphere. Rotation and 
scaling were applied to the data to align with the MDI magnetograms. The IVM 
observations were made during a 28 minute interval on 2000 June 5, for the 
sunspot in AR9026, and 2000 June 28, for the sunspot in AR9057. 

It is desirable to investigate the acoustic properties of the magnetic
field as functions of field strength and inclination. However,
in sunspot penumbrae these quantities are not easily separable since the magnetic 
field strength exhibits a close nearly linear relationship with the inclination from 
vertical (see Fig. 4 of Schunker et al. 2005).   As in Schunker et al. (2005),  
the penumbrae are divided into three roughly 
even bins, of $\gamma < 42^\circ$, ${42^\circ < \gamma < 66^\circ}$ and $\gamma > 
66^\circ$. This corresponds to a mean magnetic field strength of 1900, 1400 and 
600 Gauss for the sunspot in AR9026, and 1700, 1000 and 600 Gauss for the 
sunspot in AR9057, respectively. 

\begin{figure}
\includegraphics[trim = -20mm -20mm -20mm -5mm, clip, width=80mm]{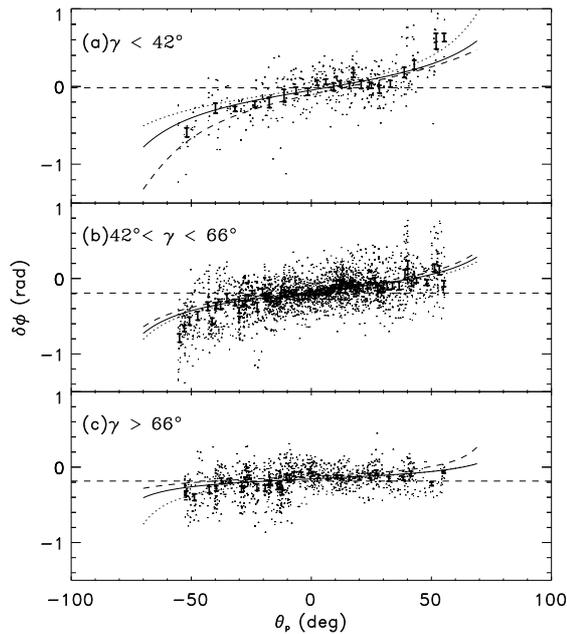}
\caption{3 mHz ingression correlation phase ($\delta \phi$) in the penumbra of 
the sunspot in AR9057 plotted against projected angle $\theta_{\rm p}$ at different 
values of magnetic field inclination. The top panel (a) shows $\gamma < 
42^\circ$, where the mean field strength is $\langle \vec{B} \rangle = 1700$ 
G, the middle panel (b) shows ${42^\circ < \gamma < 66^\circ}$, where $\langle 
\vec{B} \rangle = 1000$ G, and the bottom panel (c) shows  $\gamma > 
66^\circ$, where $\langle \vec{B} \rangle = 600$ G.  The horizontal dashed 
lines indicate the mean value of $\delta \phi$ for each panel. The error bars 
indicate the standard deviation of the mean over bins of 20 measurements in 
$\theta_{\rm p}$. The solid line is a fit for all of the displayed data; the dotted 
line is a fit for the data from 2000 June 24--28; the dashed line is a fit for 
data from 2000 June 29--2.}
\label{phi9057_3}
\end{figure}

\begin{figure}
\includegraphics[trim = -20mm -20mm -20mm -5mm, clip,width=80mm]{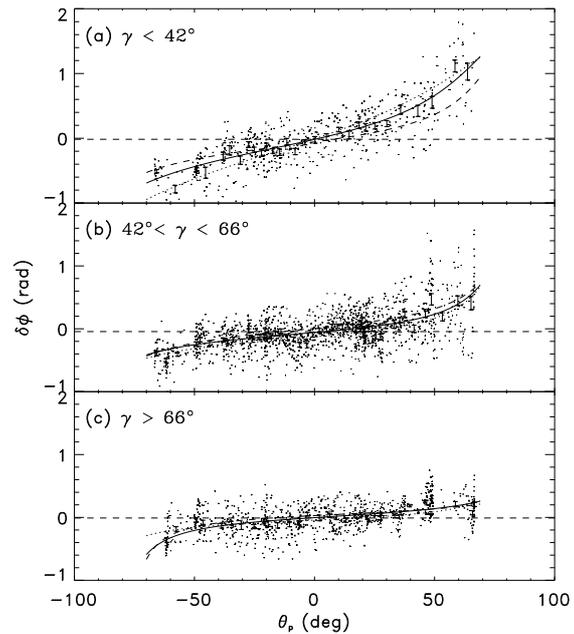}
\caption{3 mHz ingression correlation  phase ($\delta \phi$) in the penumbra of 
the sunspot in AR9026 plotted against projected angle $\theta_{\rm p}$ in three bins 
of magnetic field inclinations.  The top panel (a) shows $\gamma < 42^\circ$, 
where the mean field strength is $\langle \vec{B} \rangle = 1900$ G, the 
middle panel (b) shows $42^\circ < \gamma < 66^\circ$, where $\langle 
\vec{B} \rangle = 1400$ G, and the bottom panel (c) shows  $\gamma > 
66^\circ$, where $\langle \vec{B} \rangle = 600$ G. The horizontal dashed 
lines indicate the mean value of $\delta \phi$ for each panel. The error bars 
indicate the standard deviation of the mean over bins of 20 measurements in 
$\theta_{\rm p}$. The solid line is a fit for all of the displayed data; the dotted 
line is a fit for the data from 2000 June 3--7; the dashed line is a fit for 
data from 2000 June 8--12.}
\label{phi9026_3}
\end{figure}

\begin{figure}
\includegraphics[trim = -20mm -20mm -20mm -5mm, clip,width=80mm]{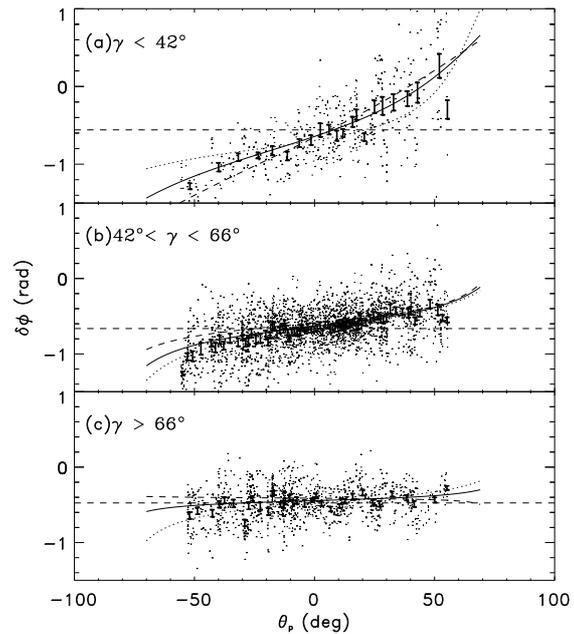}
\caption{The same as for Fig. \ref{phi9057_3}, except at 4 mHz.}
\label{phi9057_4}
\end{figure}

\begin{figure}
\includegraphics[trim = -20mm -20mm -20mm -5mm, clip,width=80mm]{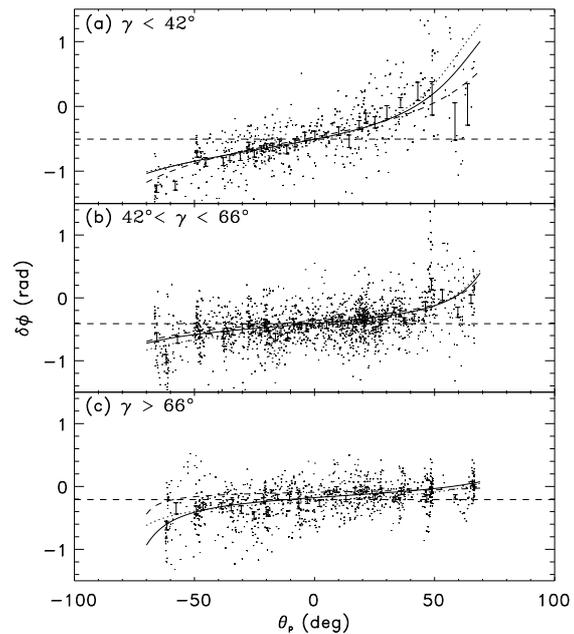}
\caption{The same as for Fig. \ref{phi9026_3}, except at 4 mHz.}
\label{phi9026_4}
\end{figure}

\begin{figure}[!t]
\includegraphics[trim = -20mm -20mm -20mm -5mm, clip,width=80mm]{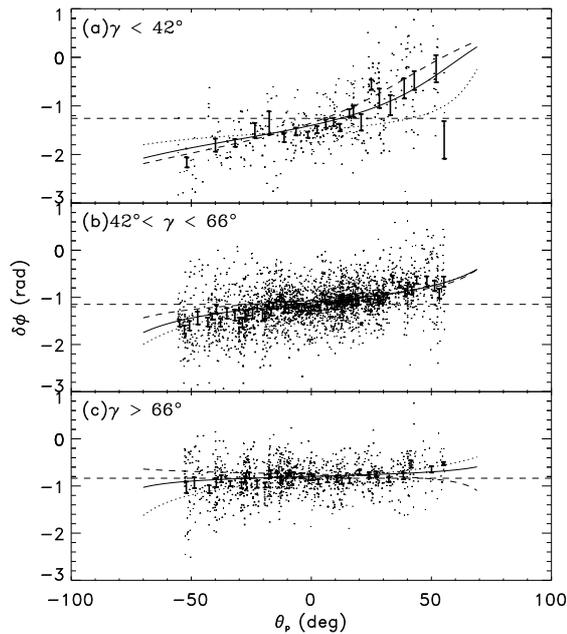}
\caption{The same as for Fig. \ref{phi9057_3}, except at 5 mHz.}
\label{phi9057_5}
\end{figure}

\begin{figure}[!t]
\includegraphics[trim = -20mm -20mm -20mm -5mm, clip,width=80mm]{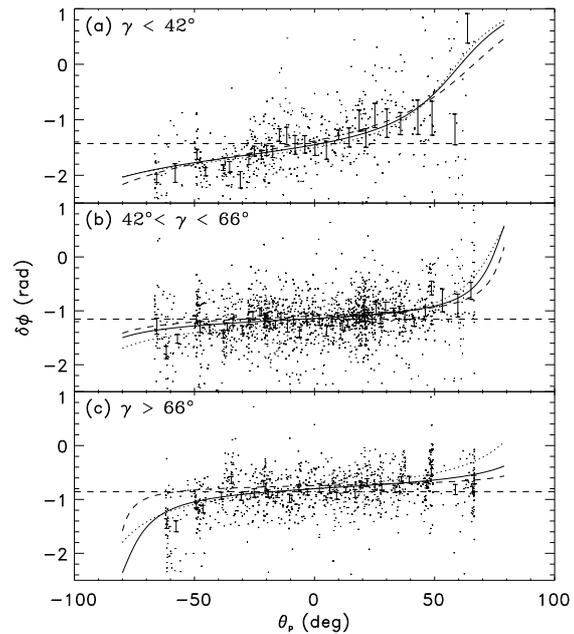}
\caption{The same as for Fig.\ref{phi9026_3} except at 5mHz.}
\label{phi9026_5}
\end{figure}

The behaviour of the ingression correlation phase with respect to 
the projected line-of-sight for both sunspots, and several
frequency bandpasses, are presented in Figs. \ref{phi9057_3} to 
\ref{phi9026_5}.  The last figure 
corresponds to Fig. 5 of Schunker et al. (2005), which showed only the results for
one sunspot with waves averaged over a 1 mHz frequency bandpass centered
at 5 mHz.  The results for two sunspots averaged over a 1 mHz frequency bandpass centered 
at  3, 4 and 5 mHz demonstrates the consistency of the results for different
spots and frequencies.

The horizontal dashed line is the average of 
$\delta \phi$ over the entire range of $\theta_{\rm p}$. 
The error bars indicate the standard deviation of the mean over 
bins of 20 measurements in $\theta_{\rm p}$. 
The observed variation of $\delta \phi$ with the projected line-of-sight angle is
consistent with the conversion of vertically propagating acoustic waves to
elliptical motion in the inclined magnetic field (Schunker et al. 2005). To quantify
this,  the variation of phase shift and amplitude (not shown) of the 
local ingression control correlations assuming elliptical motion, with the eccentricity,
inclination and amplitude of the ellipses as free parameters in the fits is modelled. 
The models will be discussed in further detail elsewhere (see also Schunker
2006), 
but the results of the fits to $\delta \phi$ are shown here to guide the eye.
The solid line is a fit for all the 
displayed data, the dotted line is a fit for data straddling the magnetogram 
(2000 June 3--7 for the sunspot in AR9026, and 2000 June 24--28 for the sunspot 
in AR9057) and the dashed line is the fit for the remaining data. These 
alternate fits are shown to assess the consistency of the results
over different time periods, since  only a single set of
IVM observations are used and it is assumed that the magnetic field is virtually 
static in time. The trends show similar properties for all time periods of the 
data.

The phase of the ingression correlation is seen to vary across the
line-of-sight angle, from $\theta_{\rm p}=-60^\circ$  to $\theta_{\rm p}=+60^\circ$, especially
at stronger fields (or smaller inclinations). This variation progressively 
decreases at weaker fields. At stronger fields where the effect is most
prominent, a slight dependency on frequency is observed in the two spots. 
At 3 mHz, $\delta \phi$ increases by about $70^\circ$  as $\theta_{\rm p}$ increases from
$-60^\circ$  and $+60^\circ$. At 5mHz, the increase in $\delta \phi$ is about 
$110^\circ$ as determined by the fits.

\section{Quiet-Sun control experiment}

The Evershed effect is a steady outflow of ${\sim2}$ km\,s$^{-1}$ seen to occur  
along the penumbral fibrils of sunspots (Evershed 1909). Since this penumbral flow
is roughly axisymmetric, its line-of-sight component 
may have a similar spatial 
dependence as the ingression correlation phase. For example, both quantities
switch sign between the side of the penumbrae extending toward disk center and the
side extending away from disk center.
The question naturally arises as to whether the variation of phase may have
non magnetic cause, and is perhaps related 
to the observed line-of-sight component of the Evershed flow.
In fact, the Evershed flows increase with radial distance from the 
sunspot centre. The effect presented here, actually decreases with radial 
distance, and is most predominant close to the umbra. In addition, we
do not expect a correlation between $\delta \phi$ and the line-of-sight
velocity, since the ingression is computed over a full annular pupil
and should not be sensitive (to first order) to solar flows. Nevertheless,
the possibility of some unknown physical relationship or instrumental, 
measurement, or analysis artifact should be tested.

\begin{figure*}
\includegraphics[width=120mm]{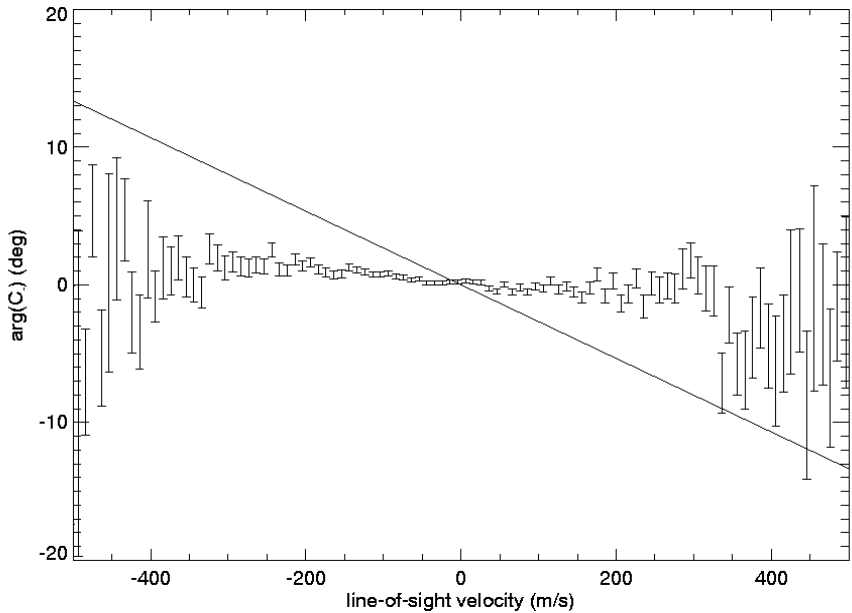}
\vspace{0.4cm}
\caption{The 5 mHz ingression phase as a function of line-of-sight velocities in the quiet sun. The diagonal solid line represents the expected supergranular flow.}
\label{ever}
\end{figure*}

A control experiment was performed to assess any dependence of the 
ingression control correlation phases in the quiet Sun with the line-of-sight
component of supergranulation, as determined from averages of full disk
MDI Dopplergrams.  This analysis must be done 
with care to remove all magnetic, including network, 
regions, which are characterized by downflows 
and may bias the result.  In addition, large-scale spatial trends must be
removed from the control correlation phases and the Dopplergrams, since
only possible correlations over the size of a typical penumbrae are
relevant to interpretations of the sunspot results.
Figure \ref{ever} shows the resulting averaged ingression control 
correlation phase with line-of-sight velocity over a wide area of the solar disk. 
The solid diagonal line represents the 
expected phase shifts of supergranular flow if the hypothesis has merit. 
This assumes that the relationship 
between the phase shifts and flows is linear and must extrapolate to values
consistent with what is observed within the two sunspots. 
The quiet-Sun phase shifts are clearly not consistent with this
expectation.  We believe that the Evershed effect 
does not significantly contribute to the effect described here and in 
Schunker et al. (2005).

\section{Discussion and conclusions}

The preliminary results first presented by Schunker et al. (2005) deserve confirmation 
and elaboration due to the implications of the results. 
Evidence for mode conversion in sunspots is critical
for our understanding of the MHD of waves and sunspots (magnetohelioseismology). 
In addition, the surface effects of the magnetic field may 
alter the results of sub-surface imaging using helioseismology within active 
regions. Here,  the results of Schunker et al. (2005) have been confirmed for a
second sunspot, and it is also shown that the results are similar
for all observed frequencies.

A frequency dependency is supported by the theory of Schunker \& Cally (2006). A wave of
lower frequency will experience the upper turning point at a lower depth than a higher frequency wave.
Therefore it will not be strongly affected by the magnetic field. In regions of
stronger field strengths, corresponding to the inner penumbrae, acoustic
waves at 5 mHz are affected more by the magnetic field as seen in the
observations presented here. However, it is not entirely clear why similar trends are
not observed in the other regions of the penumbrae.

The effect is most prevalent in strong magnetic fields.
Here, the observed variations correspond to travel time perturbations of
approximately 1 minute at all observed frequencies between 3 mHz and 5 mHz.
In comparison with travel times used to deduce sound speeds below sunspots this is
considerable. Zhao \& Kosovichev (2006)  show evidence for a similar (but smaller, 0.4 minute)
variation of travel times with azimuthal angle around a sunspot penumbra. However, their
measurements were averaged over the entire penumbrae.

As it is shown here, the 
effect is highly dependent on the field strength,
and is diminished in weaker and more inclined magnetic fields. However, 
whether this is due to the magnetic field strength or inclination is unclear, 
since in a sunspot the two properties are inseparable. Significantly,
Zhao \& Kosovichev (2006) also find a lack of variations in travel-times around the 
penumbra when they use MDI continuum intensity, rather than Dopplergrams.
We note that the interpretation of Schunker et al. (2005), namely that these variations are caused by
elliptical motion, would predict no variation with line-of-sight
angle of scalar quantities such as the continuum intensity.

The Evershed effect may be eliminated as a major cause for the effect seen in 
the ingression control correlation.  The ingression phase shift is larger 
closer to the umbra, whereas the Evershed effect is stronger close to the outer 
boundary of the penumbra. Line-of-sight supergranulation velocities do not show 
significant correlation with ingression phase shifts, which leads to the belief  
that what is being seen in penumbrae is likely a superficial 
variation in the ingression correlation phase with line-of-sight angle in magnetic
fields. 

These results have important implications for helioseismic calculations within 
active regions.  The fact that there is such a dependence of 
the phase-shift with the line-of-sight suggests that this is predominantly a 
surface effect. Zhao \& Kosovichev (2006) have argued that time-distance inversions do
not significantly change with the variation of mean line-of-sight angle of sunspots from
day to day. However, their inversions do not explicitly include or test
effects of unresolved surface terms and so cannot directly answer the
question of how the possible inclusion of those terms might change existing inferences
about subsurface conditions. 
Theory and observations of waves in active regions will certainly aid in 
understanding and ameliorating the effects of surface magnetic fields
with the goal to improve helioseismic interpretations of 
sunspot structure.

\acknowledgements
The authors would like to thank Charlie Lindsey for his valued advice and guidance.

\end{document}